\documentclass[twoside,10pt]{article}
\usepackage{fleqn,espcrc2}
\usepackage{times,mathptm}
\usepackage[dvips]{graphicx}
\newcommand{\fcaption}{%
\vspace*{-1cm}
\caption%
}
\newcommand{\beq}{\begin{equation}}
\newcommand{\eeq}{\end{equation}}
\newcommand{\bed}{\begin{displaymath}}
\newcommand{\eed}{\end{displaymath}}
\newcommand{\bea}{\begin{eqnarray}}
\newcommand{\eea}{\end{eqnarray}}
\renewcommand{\ni}{\noindent}
\renewcommand{\d}{\delta}
\renewcommand{\b}{\beta}

\newcommand{\n}{\nu}
\newcommand{\tU}{\tilde{U}}

\newcommand{\tS}{\tilde{S}}
\newcommand{\m}{\mu}
\newcommand{\s}{\sigma}

\renewcommand{\P}{{\cal P}}
\newcommand{\oh}{\frac{1}{2}}

\newcommand{\dg}{\dagger}
\newcommand{\non}{\nonumber}

\newcommand{\rf}[1]{(\ref{#1})}

\newcommand{\AmS}{{\protect\the\textfont2
  A\kern-.1667em\lower.5ex\hbox{M}\kern-.125emS}}

\title{Center Vortices at Strong Couplings}

\author{M. Faber\address{Inst. f{\"u}r Kernphysik, Tech. Univ.
Wien, A-1040 Vienna, Austria}, J. Greensite\address{Physics and Astronomy
Dept., San Francisco State Univ., San Francisco CA 94132 USA}
\thanks{Talk presented by J. Greensite.  Work supported by 
the U.S. Department 
of Energy under Grant No. DE-FG03-92ER40711.},
\v{S}. Olejn{\'\i}k\address{Inst. of Phys., Slovak Acad. of Sci., 
SK-842 28 Bratislava, Slovakia} }

\begin{document}
 
\begin{abstract}

   A long-range effective action is derived for strong-coupling
lattice SU(2) gauge theory in $D=3$ dimensions.  It is shown that center
vortices emerge as stable saddlepoints of this action.

\end{abstract}

\maketitle

   The behavior of Wilson loops $W_r(C)$ in higher group
representations $r$ is an important clue about the nature of the
confining force.  It is well known that the asymptotic
string tension $\s_r$, in
SU($N$) gauge theory, depends only on the $N$-ality of the representation
$r$.  This follows from a simple energetics argument based on the
color screening of an external charge by gluons.  On the other hand,
$W_r(C)$ is a vacuum expectation value, and it therefore carries
information about the probability distribution of vacuum fluctuations
in the \emph{absence} of external sources.  Since it is generally believed
that Wilson loops are disordered via the positional fluctuations of
some class of large-scale confining configurations, it follows that
such configurations must have the very non-trivial property that
$N$-ality=0 loops are somehow not disordered, and the asymptotic string
tension induced in other loops depends only on their $N$-ality.  The only
confining configurations known to have this property are center
vortices.  Confining instanton models, for example, tend to disorder
all loops, regardless of $N$-ality, while monopole Coulomb gas and dual
superconductor models typically predict confinement of certain (double
electric) charges of $N$-ality=0.  These properties are ruled out by general
arguments (screening) and by numerical data \cite{j3}.

  However, if the $N$-ality dependence of the string tension implies a
center vortex confinement mechanism, it follows that confinement 
in a supposedly very well-understood case, namely, strong-coupling lattice
gauge theory in $D>2$ dimensions, is also due to a vortex mechanism.
In $D>2$ dimensions, one can easily demonstrate from a strong-coupling
expansion that, e.g., $N$-ality=0 loops
have an asymptotic perimeter-law falloff.  One can, in fact, go further,
and show at strong couplings that confining disorder in $D>2$ dimension 
is due to to center disorder, rather than plaquette disorder \cite{j2}:
Consider a very large planar loop $C$, and let its minimal area be subdivided
into many sub-areas bounded by large loops $C_i$.  The question is whether
the holonomies $U(C_i)$ are uncorrelated in $D>2$ dimensions, as they
are in $D=2$ dimensions.  The test is whether, in general,
\beq
      \langle \prod_i F[U(C_i)]\rangle \stackrel{?}{=}  
          \prod_i \langle F[U(C_i)] \rangle 
\eeq
where $F[g]$ is any class function with $\int dg F[g] = 0$.  The equality
is satisfied in $D=2$ dimensions, but in any higher dimension it can be
shown that 
\beq
  \langle \prod_i F[U(C_i)] \rangle \gg  \prod_i \langle F[U(C_i)] \rangle
\eeq
which means that the holonomies $U(C_i)$ are in fact correlated in some way.  
On the other hand, if $z[g]\in Z_N$ is 
the center element closest to $g$ on the group manifold, one finds  
\beq
   \langle \prod_i z[U(C_i)] \rangle \approx
       \prod_i \langle z[U(C_i)] \rangle 
\eeq
to leading order in $\b$.  In other words, the coset elements in $SU(N)/Z_N$
associated with $U(C_i)$ holonomies are correlated; it is the center elements 
$z[U(C_i)]\in Z_N$ which are (nearly) uncorrelated \cite{j2}, and which
give rise to the area law falloff. 

\begin{figure}[t!]
\centerline{\includegraphics[width=1.0\linewidth]{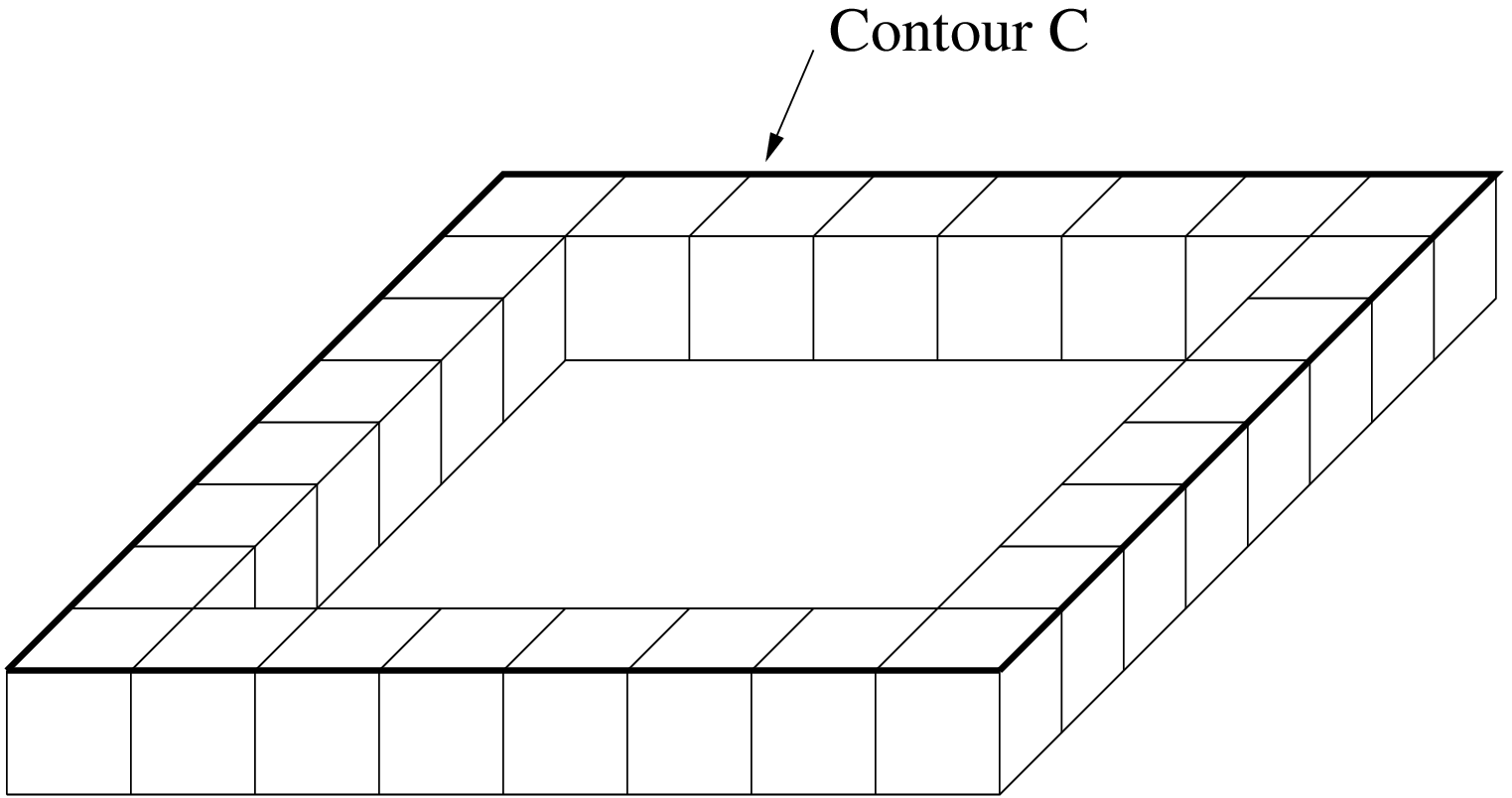}}
\fcaption{Arrangement of $U$-plaquettes in a tube around a
rectangular contour $C$ on the coarse $V$-lattice.}
\label{fig0}
\end{figure}

\begin{figure}[t!]
\centerline{\includegraphics[width=1.0\linewidth]{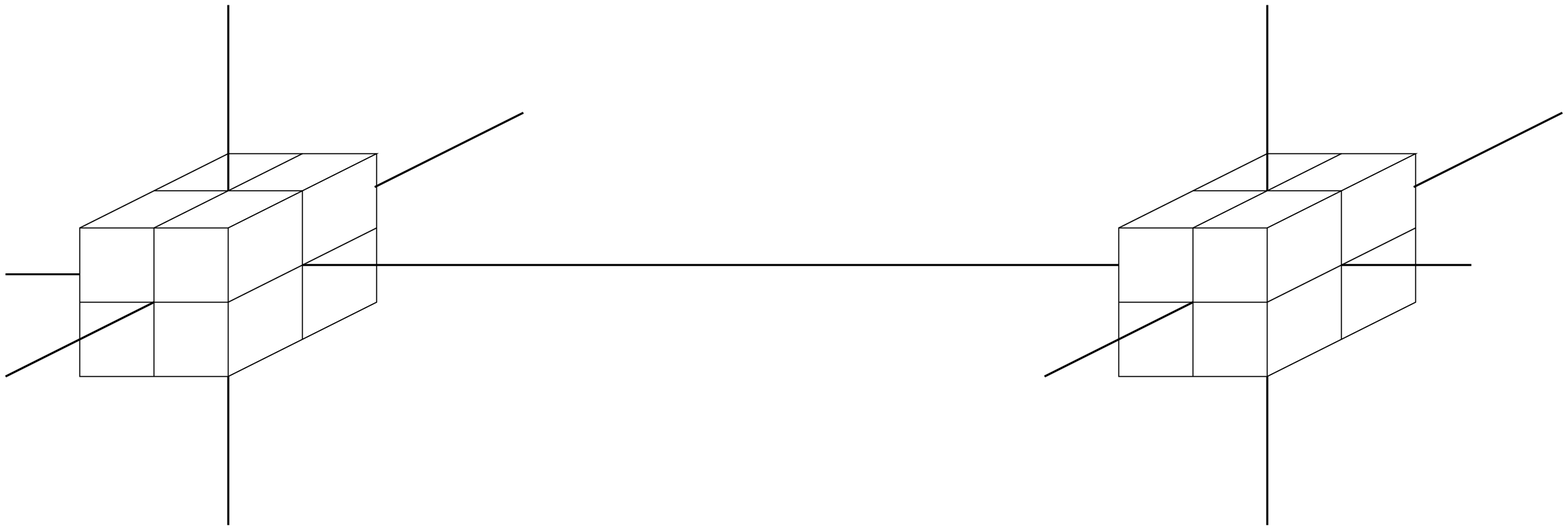}}
\fcaption{The degrees of freedom in $\tS_L[V,\tU]$.}
\label{fig1}
\end{figure}

   The fact that confining disorder is essentially 
center disorder provides some evidence
in favor of a vortex mechanism in strong-coupling lattice gauge theory.  
But we would still
like to understand more explicitly how vortices arise in this theory.  Although
center vortices are certainly not stable saddlepoints of the Wilson action,
it is possible that they might be stable saddlepoints of some long-range
effective action. Vortex structure in the vacuum should become apparent at
length scales greater than the vortex thickness, and this thickness is
expected \cite{Cas} to be on the order of the adjoint string-breaking
length, which is approximately four lattice units for $\b/4 \ll 1$.
With this in mind, we define an effective action $S_{eff}$ on a coarse
lattice, derived by integrating out links on a finer lattice, via
\bea
  \lefteqn{\exp\Bigl[S_{eff}[V]\Bigr] \equiv}
\non \\
  & & \int DU \prod_{l'} 
\d\Bigl[V^\dg_{l'}(UU..U)_{l'})-I\Bigr] e^{S_W[U]}
\label{Seff}
\eea
The length of links on the coarse $V$-lattice is equal to $L$ lattice
spacings on the original $U$-lattice.
It is not hard to see that $S_{eff}$ is non-local in $D>2$ dimensions.
For SU(2) gauge theory, $S_{eff}$ will contain large loops in all
$j=$ integer representations with only perimeter-falloff coefficients.
These terms are derived from ``tube'' diagrams, as shown in 
Fig.\ \ref{fig0}, leading to contributions such as
\bea
  \lefteqn{ \exp\Bigl[S_{eff}[V]\Bigr] 
  \supset  \int DU_{l\in C}
\prod_{l'\in C} \d\Bigl[V^\dg_{l'}(UU..U)_{l'})-I\Bigr] }
\non \\
 & &     \times
\left({\b \over 4}\right)^{4(P(C)-4)} \Bigl(\mbox{Tr}[U(C)]\Bigr)^2
\non \\
 & &      \supset 
\left( {\b\over 4}\right)^{4(P(C)-4)} \chi_1[V(C)]
\eea
We would like to derive, instead, an effective action where the 
leading contributions
to any Wilson loop on the $V$-lattice (including $j=$ integer loops)
are obtained from \emph{local}
terms.  To achieve this, working in $D=3$ dimensions,
we consider integrating (in eq.\ \rf{Seff})
all links on the the $U$-lattice except those belonging
to cubes of volume $2^3$ centered at sites on the $V$-lattice, as shown
in Fig.\ \ref{fig1}.
This procedure defines the effective action $\tS_L$
\bea
   Z &=&
   \int DV \int \prod_{l\in 2-cubes} d\tU_l 
   \int \prod_{l''\not\in 2-cubes} dU_{l''} 
\non \\
 & & \prod_{l'} \d\Bigl[V^\dg_{l'}(UU..U)_{l'})-I\Bigr] e^{S_W[U]} 
\non \\
   &\equiv& \int DV \int \prod_{l\in 2-cubes} d\tU_l ~ 
                      \exp\Bigl[\tS_L[V,\tU]\Bigr]
\eea

\begin{figure}[t!]
\centerline{\includegraphics[width=1.0\linewidth]{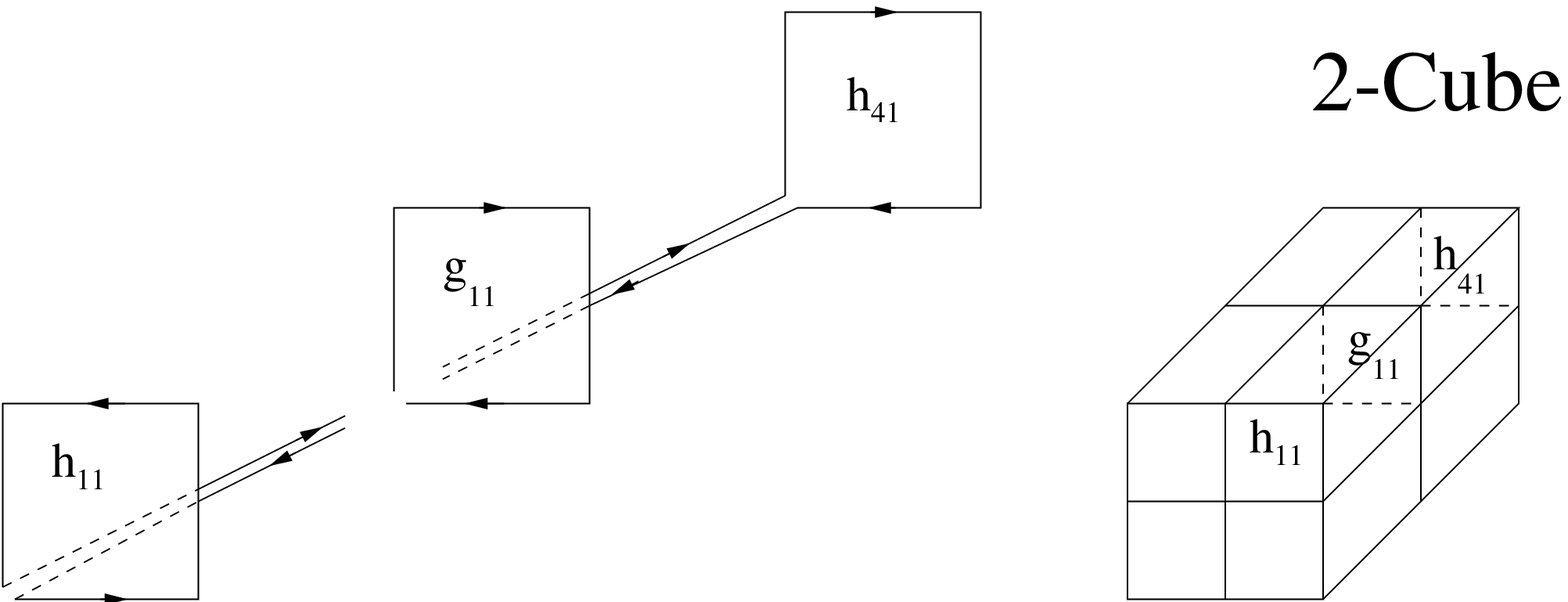}}
\fcaption{Three plaquette variables on the 2-cube.}
\label{fig4}
\end{figure}

   Introduce group-valued plaquette variables $h,g$ lying in the
2-cubes centered at $V$-lattice sites, with $h$ variables running around
plaquettes on the surface of the 2-cube, and $g$ variables running around
plaquettes in the interior, as indicated in Fig.\ \ref{fig4}. 
After integrating over $U$-links not in 2-cubes, we find
\bea
   Z &\approx& 
\int DV D\tU ~ \exp\left[ {\b\over 2} \sum(\mbox{Tr}[h] 
                         + \mbox{Tr}[g]) \right. 
\non \\
  & &  + 2\left(\b\over 4\right)^{4(L-2)}\sum_{l'}
  f_{l'}^{ijkl} \mbox{Tr}[h^\dg_{ij}V_{l'}
            h^\dg_{kl}V^\dg_{l'}]
\non \\
   & & \left.  + 2\left({\b\over 4}\right)^{L^2-4} \sum_{P'}
              \mbox{Tr}[VgVgV^\dg g^\dg V^\dg g^\dg] \right]
\eea
Now change integration variables from $\tU$ links to plaquette variables
$h,g$ on 2-cubes.  This introduces a Bianchi constraint on each 1-cube
\cite{Batrouni}.  Symbolically,
\bea
     \d[\mbox{Bianchi}] &=& \d[hghghg - I]
\non \\
                   & =& \sum_{j=0,\oh,1,..} (2j+1) \chi_j[hghghg]
\eea
After integrating over the $g$-variables, we obtain (writing only terms
of low order in $\b$ and $h$)
\bea
   Z &\approx&
   \int DV Dh \prod_{2-cubes~K}
\non \\
  & & \left\{ 1  + 2\left({\b\over 4}\right)^3 \sum_{c\in K} \right. 
\chi_\oh[(hhh)_c] 
\non \\
      & &\left.  + 2\left({\b\over 4}\right)^4 
                 \sum_{\stackrel{adjacent}{c_1c_2\in K}} 
                  \chi_\oh[(hhh)_{c_1}(hhh)_{c_2}] + ... \right\}
\non \\
  &\times& \exp\left[ {\b\over 2} \sum \mbox{Tr}[h] \right. 
\non \\ 
  & &  + 2\left(\b\over 4\right)^{4(L-2)}\sum_{l'}
  f_{l'}^{ijkl} \mbox{Tr}[h^\dg_{ij}V_{l'}
            h^\dg_{kl}V^\dg_{l'}] 
\non \\
  & & \left.  + 2\left({\b\over 4}\right)^{L^2} \sum_{P'}
              \mbox{Tr}[V V V^\dg  V^\dg ] \right]
\non \\
   &\approx& \int DV Dh ~ \exp\Bigl[S_L[V,h]\Bigr]
\eea
\bigskip

   This resembles an adjoint-Higgs theory, with an SU(2)
gauge field $V_\m$ coupled to 24 ``matter'' fields $h$ in the
adjoint representation.  Note that for coarse lattices (large $L$), 
the ``Higgs'' potential term is much larger than the ``kinetic''
and pure-gauge
($V$-plaquette) terms, so the $h$-fields fluctuate almost independent
of $V_\m$.

   Next we fix some of the $h$-fields by a choice of unitary gauge,
leaving a remnant $Z_2$ gauge invariance, and integrate out the remaining
$h$ d.o.f.\ to find
\bea
    \lefteqn{  S_{eff}[V] \approx 
S_{link}[V,\langle h \rangle_h] + S_{plaq}[V] }
\non \\
   && = 2\left({\b\over 4}\right)^{4(L-2)}
           \sum_{l'} f_{l'}^{ijkl}\mbox{Tr} \Bigl[
      \langle h^\dg_{ij} \rangle_h V_{l'} 
      \langle h^\dg_{kl} \rangle_h V^\dg_{l'} \Bigr]
\non \\
       &&  + 2\left({\b\over 4}\right)^{L^2}
             \sum_{P'} \mbox{Tr}[VVV^\dg V^\dg]
\eea
where the expectation value of the
plaquette variable $h_{ij}$ is determined almost entirely, at large
$L$ and small $\b$, by the gauge-fixed ``Higgs potential'' term 
alone (c.f.\ \cite{strong} for details).

  We are now ready to look for saddlepoints of this effective
action.  We find that $S_{link}$ is maximized at
\bea
         V_\m(\vec{n}) & =&  Z_\m(\vec{n}) 
                       \times g(\vec{n}) g^\dg(\vec{n}+\m)
\non \\
   Z_\m    & =&  \pm 1
\eea
where $ g(\vec{n}) g^\dg(\vec{n}+\m)$ is { fixed} by the particular 
unitary gauge
choice, while $ S_{plaq}$ is maximized if $ ZZZZ=+1$.  This is the
unitary gauge ground state.

   Consider a thin center vortex created on this state by a discontinuous 
gauge transformation, e.g.\
\bea
      Z_y(\vec{n}) & =&  \left\{ \begin{array}{rl}
                    -1 & n_1 \ge 2, ~ n_2=1 \cr 
                    +1 & \mbox{otherwise} \end{array} \right.
\non \\
     Z_x(\vec{n}) & =&  Z_z(\vec{n}) = 1 
\eea
         
  This configuration is { stationary}:  $ S_{link}[V]$ is
still a maximum, $ S_{plaq}$ is extremal (max or min) on all plaquettes.
Stability depends on the eigenvalues of
\bea
    \lefteqn{ {\d^2 S_{eff} \over \d V_\m(n_1) \d V_\n(n_2)} = }
\non \\
   & &     {\d^2 S_{link} \over \d V_\m(n_1) \d V_\n(n_2)}
      + {\d^2 S_{plaq}  \over \d V_\m(n_1) \d V_\n(n_2)} 
\eea
and we find
\bea
      {\d^2 S_{link}  \over  \d V_\m(n_1) \d V_\n(n_2)}
         &\sim& \left({\b\over 4}\right)^{4(L-2)+12}
\non \\
      {\d^2 S_{plaq}  \over  \d V_\m(n_1) \d V_\n(n_2)}
         &\sim&  \left({\b\over 4}\right)^{L^2}
\eea

   The crucial observation is that for $\b/4 \ll 1$ and
\beq
      4(L-2) + 12 < L^2  ~~~\Longrightarrow~~~ L \ge 5
\eeq
the contribution of $ \d^2 S_{plaq}/\d V \d V$ to the stability 
matrix (and therefore to the eigenvalues of the stability matrix) 
is negligible compared to
$ \d^2 S_{link}/\d V \d V$, which implies 

\begin{itemize} 
\item {\bf Vortex Stability:~}
  The thin vortex is a stable saddlepoint of the 
full effective action $S_{eff}$ at $L\ge 5$.
\item {\bf Vortex Thickness:~}
   A ``thin'' vortex on the $V$-lattice means thickness less than $L$ on the
$U$-lattice.  This means that stable 
center vortices are $\approx 4-5$
lattice spacings thick.  As it happens, this is also the distance where the 
adjoint string breaks.
The correspondence between the adjoint string-breaking scale, and
the thickness of center vortices, was suggested by us some years ago,
in ref.\ \cite{Cas}, in connection with Casimir scaling of adjoint loops.
\item {\bf Percolation:~}
   From $S_{eff}$, we see that center vortices in D=3 dimensions 
cost, in action,
\beq
       8 \left( {\b\over 4} \right)^{L^2} / \mbox{unit length}
\eeq
while the entropy is O(1)/unit length. Since vortex entropy exceeds
vortex action, these configurations percolate through the
lattice, and confine $N$-ality $\ne 0$ charge.
\end{itemize}

   We conclude that even in strong-coupling lattice gauge theory, the
asymptotic string tension is due to the disordering effect of center vortices.
Details of the analysis presented here can be found in
ref.\ \cite{strong}.

\end{document}